\documentclass[12pt]{article}
\usepackage{a4,epsfig,amsmath,amssymb,overcite}
\begin{document}
\begin{center}
{\large\bf Sensitivity of Phage Lambda upon Variations of the Gibbs Free Energy}\\
  \vspace*{1cm}
  \centering{Audun Bakk\footnote{NORDITA, Blegdamsvej 17, DK-2100 Copenhagen, Denmark}\renewcommand{\thefootnote}{\fnsymbol{footnote}}\footnote{E-mail: audunba@nordita.dk}, Ralf Metzler$^1$\footnote{Corresponding author. E-mail: metz@nordita.dk}, and Kim Sneppen$^1$\footnote{E-mail: sneppen@nordita.dk}\\
}
  \vspace*{0.5cm}
  \centering{(\today)}
\end{center}

\vspace{0.5 cm}
\noindent
{\bf Short title:} Sensitivity of phage lambda\\

\noindent
{\bf Key-words:} bacteriophage lambda, repressor, sensitivity, genetic networks, genetic regulation
\begin{abstract}
We investigate the sensitivity of production rates (activities) of the regulatory proteins CI (repressor) and Cro at the right operator ($O_{\rm R}$) of bacteriophage lambda. The DNA binding energies of CI, Cro, and RNA polymerase are perturbed to check the uncertainty of the activity, due to the experimental error, by means of a computational scattering method according to which the binding energies are simultaneously chosen at random around the literature values, with a width corresponding to the experimental error. In a grand canonical ensemble, with the randomly drawn protein-DNA binding energies as input, we calculate the corresponding activities of the promoters $P_{\rm RM}$ and $P_{\rm R}$. By repeating this procedure we obtain a mean value of the activity that roughly corresponds to wild-type (unperturbed) activity. The standard deviation emerging from this scheme, a measure of the sensitivity due to experimental error, is significant (typically $>$ 20\% relative to wild-type activity), but still the promoter activities are sufficiently separated to make the switch feasible. We also suggest a new compact way of presenting repressor and Cro data.   

\end{abstract}

\vspace{1cm}\noindent
{\sf Dedicated to Joshua Jortner on the occasion of his 70$^{\rm th}$ birthday.}

\renewcommand{\thefootnote}{\fnsymbol{footnote}}
\section*{Introduction}
The situation is simple: we know the genes, but we do not know how they are regulated or transcribed precisely. To understand how genetic networks behave appears a major challenge in the ``post genomic'' era\cite{Rao:01}. An example of a class of small genetic networks, often suitable for theoretical modeling, are the so-called genetic switches. Shortly explained, a genetic (regulatory) switch is a system consisting of a DNA region (operator) and regulatory protein(s) that are able to bind to this operator in order to foster or inhibit the transcription of a certain gene of the DNA.\cite{Snustad:00} Several genetic switching systems have been studied extensively, e.g., the tryptophan repressor and the {\it lac}Operon in {\it E.\ coli} (procaryotic systems),\cite{Alberts:94} and regulation of the {\it gal} genes in yeast\cite{Ptashne:02} (eucaryotic systems).\footnote{Nomenclature: genes are denoted with italicized letters and their protein products with Roman letters (first letter capitalized).}

In this work we want to study the sensitivity upon variations of the protein-DNA binding energies of the right operator ($O_{\rm R}$) of bacteriophage lambda (phage $\lambda$) in respect to experimental error. This operator is in general described elsewhere, e.g., by Ptashne\cite{Ptashne:92}. In brief, $O_{\rm R}$ is regulating two important genes to either side; {\it c}I and {\it cro} which in turn act as a template for the regulatory proteins CI and Cro, respectively. Upon injection of DNA from phage $\lambda$ into an {\it E. coli} bacterium, $O_{\rm R}$ is crucially important to decide the fate of the bacterium. I.e., the switch funnels entry into the dormant lysogenic state, or into the lytic state leading to the formation of new $\lambda$-phages by help of the facilities of the {\it E. coli} cell, and, ultimately to the death of the {\it E. coli} cell. Partially overlapping the switch are the promoter regions $P_{\rm RM}$, that initiates {\it c}I transcription, and  $P_{\rm R}$, that initiates {\it cro} transcription. 

We present a new method for analyzing the sensitivity of the activity at the two promoters of $O_{\rm R}$, taking into account the experimental error in the experiments used to determine the Gibbs free energies (GFEs) of the regulatory proteins and RNA polymerase (RNAP), by simultaneous random perturbations of the GFEs. A new way of presenting repressor data, where Cro data is implicitly given, is also discussed. 

\section*{Modeling the system}
A fundamental assumption in this work is the widely accepted view that the protein-DNA binding/unbinding rates of CI, Cro, and RNAP are in equilibrium, i.e., protein associations with DNA are much faster (fractions of a second) compared with relevant time-scales for protein production and thus activity (seconds)\cite{Ackers:82,Shea:85,Aurell:02}. In equilibrium, the protein-DNA associations of CI dimers (CI$_2$), Cro dimers (Cro$_2$), and RNAP to $O_{\rm R}$ of phage $\lambda$ occur in, presently identified, 40 experimentally distinguishable states. The associated probability $f_s$ for finding the system in one of the 40 states $s$ is\cite{Hill:60,Ackers:82} 
\begin{equation}
   \label{eq:f_s}
   f_s=\frac{\exp{\boldsymbol{(}-\Delta G(s)/(RT)\boldsymbol{)}}\,[{\rm CI}_2]^{i_s}\,[{\rm Cro}_2]^{j_s}\,[{\rm RNAP}]^{k_s}}{\sum_s\exp{\boldsymbol{(}-\Delta G(s)/(RT)\boldsymbol{)}}\,[{\rm CI}_2]^{i_s}\,[{\rm Cro}_2]^{j_s}\,[{\rm RNAP}]^{k_s}}\,,
\end{equation}
where $R=8.31$ J/(mol K) is the gas constant, $T=310$ K is the absolute temperature ($37^{\circ}$C), and $\Delta G(s)$ is the GFE difference between state $s$ and the unoccupied state, i.e., protein-DNA binding energy. All concentrations ([X]) refer to the unbound state in solution. $i_s$, $j_s$, and $k_s$ are the numbers of  CI dimers, Cro dimers, and RNAP bound to $O_{\rm R}$ in state $s$, respectively.

The different $\Delta G(s)$ in Eq.\ (\ref{eq:f_s}) are in general a sum of GFE originating from the individual and cooperative bindings of the proteins at the three different binding sites of $O_{\rm R}$ (for details, e.g., see Figure 1 of Shea and Ackers\cite{Shea:85}). In this work we apply GFE data of CI from Koblan and Ackers\cite{Koblan:92}, Cro data from Darling et al.\cite{Darling:00b}, and RNAP data from Ackers et al.\cite{Ackers:82}. These binding energies are summarized in Table \ref{tab:1}. 
\begin{table}[ht]
\begin{center}
\caption{\label{tab:1}Protein-DNA binding energies (GFEs) for CI from Koblan and Ackers\cite{Koblan:92}, Cro from Darling et al.\cite{Darling:00b}, and RNAP from Ackers et al.\cite{Ackers:82}. All GFEs are given in kcal/mol and limits ($\pm$) correspond to 67\% confindence intervals. $\Delta G_1$ is the GFE associated with the binding between CI and operator site $O_{\rm R}$1, etc.\ (see, e.g., Ptashne\cite{Ptashne:92} for an explanation/illustration of the different operator sites). $\Delta G_{12}$ is the GFE associated with coopertaive binding between CI at $O_{\rm R}$1 and $O_{\rm R}$2, etc. GFEs with a prime (e.g., $\Delta G_{1'}$) correspond to Cro data, otherwise analogous to CI notation. $\Delta G_{\rm RM}$ and $\Delta G_{\rm R}$ are GFEs associated with binding of RNAP to $P_{\rm RM}$ and $P_{\rm R}$, respectively. Experimental data are obtained {\it in vitro} in 200 mM KCl, resembling ``physiological'' conditions\cite{Kao-Huang:77, Ackers:82}. CI and Cro are both assumed to obey a monomer-dimer equilibrium in solution where the free energies of dimerization are $-11.0$ kcal/mol\cite{Koblan:91} and $-8.7$ kcal/mol,\cite{Darling:00a} respectively. In lack of Cro data at 37$^{\circ}$C, at which temperature CI and RNAP data are measured, these data are obtained at 20$^{\circ}$C.} 
\centering{
\begin{tabular}{lllr}
         \hline {\bf CI}   && $\Delta G_1$       & -12.5 $\pm$ 0.3\\ 
                           && $\Delta G_2$       & -10.5 $\pm$ 0.2\\                                                    && $\Delta G_3$       & -9.5 $\pm$ 0.2\\                                                     && $\Delta G_{12}$    & -2.7 $\pm$ 0.3\\                                                     && $\Delta G_{23}$    & -2.9 $\pm$ 0.5\\
         \hline {\bf Cro}  && $\Delta G_{1'}$    & -12.0 $\pm$ 0.1\\                                                    && $\Delta G_{2'}$    & -10.8 $\pm$ 0.1\\
                           && $\Delta G_{3'}$    & -13.4 $\pm$ 0.1\\                                                    && $\Delta G_{12'}$   & -1.0 $\pm$ 0.2\\                                                     && $\Delta G_{23'}$   & -0.6 $\pm$ 0.2\\                                                     && $\Delta G_{123'}$  & -0.9 $\pm$ 0.2\\
         \hline {\bf RNAP} && $\Delta G_{\rm RM}$& -11.5 $\pm$ 0.5\\                                                    && $\Delta G_{\rm R}$ & -12.5 $\pm$ 0.5\\ \hline
\end{tabular}}
\end{center}
\end{table}

\begin{table}
\begin{center}
\caption{\label{tab:2}Gibbs free energies (GFEs) of the different protein associations to $O_{\rm R}$ of phage $\lambda$ (in state $s$) of CI dimers (R),\cite{Koblan:92} Cro dimers(C),\cite{Darling:00b} and RNAP.\cite{Ackers:82} ``0'': empty site, ``$\longleftrightarrow$'': cooperative interaction, and ``Terms'': GFE terms due to Table \ref{tab:1}. GFEs are measured in kcal/mol relative to the unbound state of zero GFE (Reference state; $s=1$).}\vspace*{0.3cm}
\centering{\footnotesize\begin{tabular}{lccccclr}                                                     \hline $s$& $O_{\rm R}$3 & & $O_{\rm R}$2& &$O_{\rm R}$1 &Terms &GFE\\  
   \hline     1& 0 & & 0& &0 &Reference state &0\\                                                         2& 0 & & 0& &R  &$\Delta G_1$ &-12.5\\
              3& 0 & &R& &0 &$\Delta G_2$ &-10.5\\                                                         4& R & & 0& &0 &$\Delta G_3$ &-9.5\\
              5& 0 & & 0& &C &$\Delta G_{1'}$ &-12.0\\
              6& 0 & & C& &0 &$\Delta G_{2'}$ &-10.8\\                                                     7& C& & 0& &0 &$\Delta G_{3'}$ &-13.4\\                                                      8& RNAP& &0& &0&$\Delta G_{\rm RM}$ &-11.5\\                                                 9& 0 & & &RNAP &&$\Delta G_{\rm R}$ &-12.5\\                                                 10& 0& & R &$\longleftrightarrow$ & R
                &$\Delta G_1+\Delta G_2+\Delta G_{12}$ &-25.7\\
              11& R & & 0& &R &$\Delta G_1+\Delta G_3$ &-22.0\\                                            12& R &$\longleftrightarrow$ & R& &0                                                           &$\Delta G_2+\Delta G_3+\Delta G_{23}$ &-22.9\\
              13& 0 & & C& $\longleftrightarrow$&C                                                          &$\Delta G_{1'}+\Delta G_{2'}+\Delta G_{12'}$ &-23.8\\
              14& C& & 0& &C&$\Delta G_{1'}+\Delta G_{3'}$ &-25.4\\                                        15& C&$\longleftrightarrow$  & C& &0                                                             &$\Delta G_{2'}+\Delta G_{3'}+\Delta G_{23'}$ &-24.8\\                                   16& RNAP& & &RNAP & &$\Delta G_{\rm RM}+\Delta G_{\rm R}$ &-24.0\\                           17& 0 & & C& &R &$\Delta G_1+\Delta G_{2'}$ &-23.3\\                                         18& 0 & & R& &C&$\Delta G_{1'}+\Delta G_2$ & -22.5\\                                         19& R & & 0& &C&$\Delta G_{1'}+\Delta G_3$ &-21.5\\
              20& C& & 0& &R &$\Delta G_1+\Delta G_{3'}$ &-25.9\\                                          21& R & & C& &0 &$\Delta G_{2'}+\Delta G_3$ &-20.3\\
              22& C& & R& &0 &$\Delta G_2+\Delta G_{3'}$ &-23.9\\
              23& R & & & RNAP&&$\Delta G_{\rm R}+\Delta G_3$&-22.0\\                                      24&  RNAP& & R& &0 &$\Delta G_2+\Delta G_{\rm RM}$ &-22.0\\                                  25& RNAP&&0& &R&$\Delta G_1+\Delta G_{\rm RM}$ &-24.0\\                                      26& C& & &RNAP&&$\Delta G_{\rm R}+\Delta G_{3'}$ &-25.9\\                                    27& RNAP& & C& & 0 &$\Delta G_{2'}+\Delta G_{\rm RM}$ &-22.3\\
              28& RNAP & & 0& &C&$\Delta G_{1'}+\Delta G_{\rm RM}$&-23.5\\                                 29& R && R&$\longleftrightarrow$&R                                                             &$\Delta G_1+\Delta G_2+\Delta G_3+\Delta G_{12}$ &-35.2\\
              30& C&$\longleftrightarrow$ & C&$\longleftrightarrow$ &C                                       &$\Delta G_{1'}+\Delta G_{2'}+\Delta G_{3'}+\Delta G_{123'}$&-37.1\\
              31& C& & R&$\longleftrightarrow$ &R &                                                         $\Delta G_1+\Delta G_2+\Delta G_{3'}+\Delta G_{12}$ &-39.1\\                                   32& R & & C& &R &$\Delta G_1+\Delta G_{2'}+\Delta G_3$&-32.8\\                               33& R &$\longleftrightarrow$ & R& &C                                                           &$\Delta G_{1'}+\Delta G_2+\Delta G_3+\Delta G_{23}$&-34.9\\                               34& R & & C&$\longleftrightarrow$ &C                                                           &$\Delta G_{1'}+\Delta G_{2'}+\Delta G_3+\Delta G_{12'}$&-33.3\\                           35& C&& R& &C&$\Delta G_{1'}+\Delta G_2+\Delta G_{3'}$&-35.9\\
              36& C&$\longleftrightarrow$ & C& &R                                                            &$\Delta G_1+\Delta G_{2'}+\Delta G_{3'}+\Delta G_{23'}$&-37.3\\
              37& RNAP & & R&$\longleftrightarrow$ &R &                                                         $\Delta G_1+\Delta G_2+\Delta G_{\rm RM}+\Delta G_{12}$ &-37.2\\                        38& RNAP & & C&$\longleftrightarrow$ &C                                                       &$\Delta G_{1'}+\Delta G_{2'}+\Delta G_{\rm RM}+\Delta G_{12'}$&-35.3\\                     39& RNAP & & C& &R&$\Delta G_1+\Delta G_{2'}+\Delta G_{\rm RM}$&-34.8\\                      40& RNAP & & R& &C                                                                             &$\Delta G_{1'}+\Delta G_2+\Delta G_{\rm RM}$&-34.0\\
             \hline
\end{tabular}}
\end{center}
\end{table}

In Table \ref{tab:2} we list the corresponding 40 different states of protein-DNA associations. Throughout this work we have for simplicity assumed a constant free RNAP concentration of 30 nM\cite{Shea:85}. Note that in lack of Cro data at 37$^{\circ}$C, at which temperature CI and RNAP data are taken, the Cro data used in the following were obtained at 20$^{\circ}$C. It is assumed that the latter data provide a reasonable estimate for the process at 37$^{\circ}$C.  

The main purpose of this paper is to study the sensitivity of production rates (activities) with respect to the experimental error of the GFEs. To this end, we assume that the transcription initiation (isomerization rate) is the rate-limiting step in protein synthesis\cite{Ptashne:80,McClure:80}. Accordingly, activity will be defined as the product of isomerization rate times the probability of RNAP occupancy of the promoter. The latter probability is a sum of the $f_s$ in Eq.\ (\ref{eq:f_s}). In what follows, we use the same rate constants as Shea and Ackers in enumerating these activities\cite{Shea:85}. 

\section*{Results and discussion}
In a previous study we analyzed the sensitivity of $O_{\rm R}$ through a systematic one-by-one perturbation scheme of the GFEs, with a data set without monomer-dimer equilibrium for Cro\cite{Bakk:03}. Each individual GFE (corresponds to those in Table \ref{tab:1}) was perturbed $\pm 1$ kcal/mol, one-by-one, whereupon the change in activity compared to wild-type (unperturbed) activity was calculated. Bakk et al.\ show in this work that for a lysogen the sensitivity of the activity is low (upon CI and Cro perturbations), while this sensitivity is increasing for protein concentrations around induction where the $\lambda$-switch is turning over from the lysogenic to the lytic pathway.  

The distinct novel feature in this work is that we perform a computational scattering method, where the different GFEs are randomly chosen in the parameter hyperspace and applied in the model simultaneously. This implies that for each GFE (see Table \ref{tab:1}), we draw from a Gaussian distribution with standard deviation equal to the experimental uncertainty (indicated by $\pm$ in Table \ref{tab:1}).\footnote{67\% confidence interval corresponds to a Gaussian distribution around the mean value with standard deviation equal to the experimental uncertainty.} Then, 13 new values for the GFE are obtained and the activities at both promoters are then evaluated. This procedure is performed $10^3$ times, which we checked to be significant to ensure reliable statistics, whereupon the mean value (mean) and standard deviation (SD) are calculated from this set. The latter value will reflect typical uncertainty of the activities due to the experimental error of the GFEs. Here we define the sensitivity of the activity as the ratio between the standard deviation ensuing the computational scattering and wild-type (unperturbed) activity. 

\begin{figure}
   \centerline{\epsfig{figure=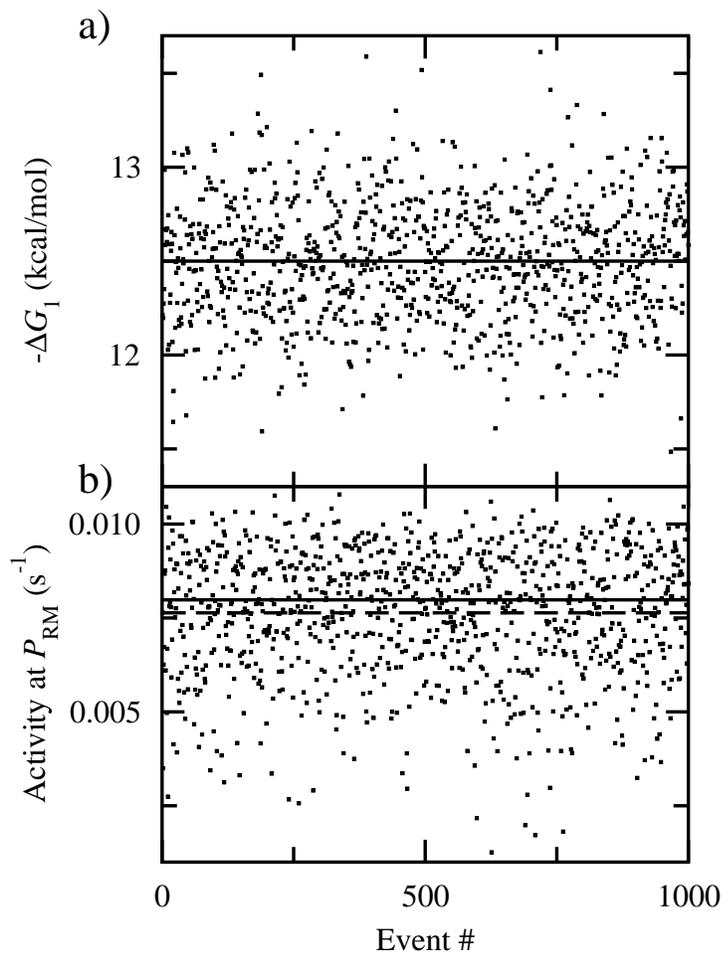,width=10cm}}
    \caption{\label{fig:1}{\bf a)} Scattering of the protein-DNA binding energy $\Delta G_1$ randomly drawn from a Gaussian distribution with mean of -12.5 kcal/mol (horizontal line) and SD of 0.3 kcal/mol (see Table \ref{tab:1}). The latter value corresponds to 67\% confidence intervals in the experiments. {\bf b)} Corresponding scattering of the activity, due to variations of all GFEs in the same run, at promoter $P_{\rm RM}$ for  [CI$_{\rm t}]=200$ nM and zero Cro concentration (typical for a lysogen). Continous horizontal line \newline({\bf ---------}) corresponds to wild type activity (0.0081 s$^{-1}$) and scattered horizontal line ({\bf - - - - -}) corresponds to the mean activity of the 1000 scattered values in this plot (0.0077 s$^{-1}$). ``Event \#'' refers to the number in the series of the randomly drawn binding energies out of 1000 realizations.}
\end{figure}

Figure \ref{fig:1}a shows how the parameter $\Delta G_1$ is scattered around the mean value -12.5 kcal/mol, with SD of 0.3 kcal/mol as given in Table \ref{tab:1}, for 1000 realizations (random draws). 
Figure \ref{fig:1}b give a corresponding example of how activity is spread due to variations of all GFEs in the same run. Note in this particular example that some of the scattered data points are shifted toward very low values leading to a mean value of the scattered activities which is lower (0.0077 s$^{-1}$) compared to the wild-type activity (0.0081 s$^{-1}$). However, as also discussed below, skewness, here and in the other simulations, is not very pronounced.    
\begin{figure}[ht]
   \centerline{\epsfig{figure=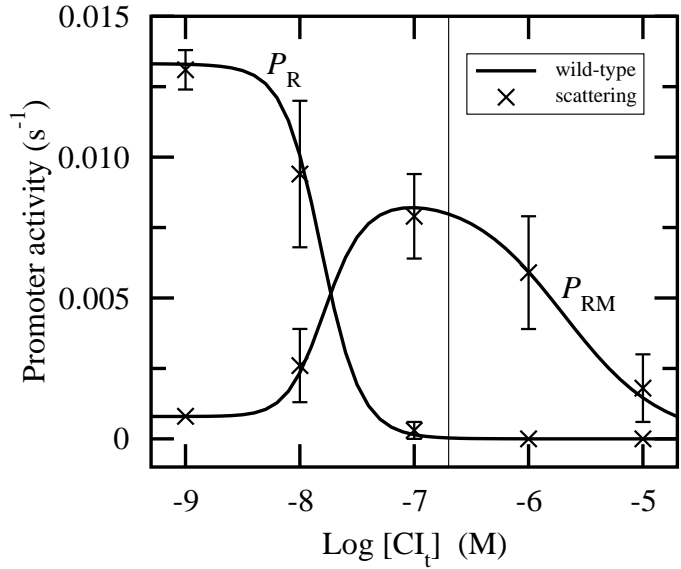,width=10cm}}
   \caption{\label{fig:2} Promoter activity versus total CI concentration for [Cro$_{\rm t}]$ $\approx 0$. $P_{\rm RM}$ corresponds to {\it c}I activity and $P_{\rm R}$ corresponds to {\it cro} activity. Fully drawn curves (``wild-type'') correspond to experimental GFE data listed in Table \ref{tab:1}, where  CI data are from Koblan and Ackers\cite{Koblan:92}, Cro data from Darling et al.\cite{Darling:00b}, and RNAP data from Ackers et al.\cite{Ackers:82}. Promoter activity corresponds to the number of RNAP-DNA complexes that becomes transcriptionally active per second. ``Scattering'' ($\boldsymbol{\times}$) are mean values of the activities obtained from the computational scattering (described in main text) associated with standard deviations (only indicated for deviations $>0.3\times 10^{-3}$ s$^{-1}$). Thin vertical line indicates lysogenic concentration ($\approx 200$ nM). Abscissa is drawn on logarithmic (decadic) scale.}
\end{figure}
The obtained mean values are very close to the wild-type values. This is not {\it a priori} obvious, because these values originate from random draws in a Gaussian distribution of the GFEs, which in turn enters exponents in the grand canonical partition function (Eq.\ (\ref{eq:f_s})) that might produce a skewness in the distribution of the activities around the mean. A general feature is that the SD relative to the wild-type activity, i.e., the sensitivity, is large and that the sensitivity is largest for a combination of moderate/large repressor concentrations and low activity. On the other hand, it is known from experiments that the robustness upon perturbations, in particular of the lysogenic state, is high\cite{Bailone:79,Little:99,Aurell:02}. Thus, in light of these latter mentioned studies, and despite the resulting large uncertainty of the activities due to the experimental error, as found here, a lysogen remains stable due to the perturbations.

\begin{figure}[ht]
   \centerline{\epsfig{figure=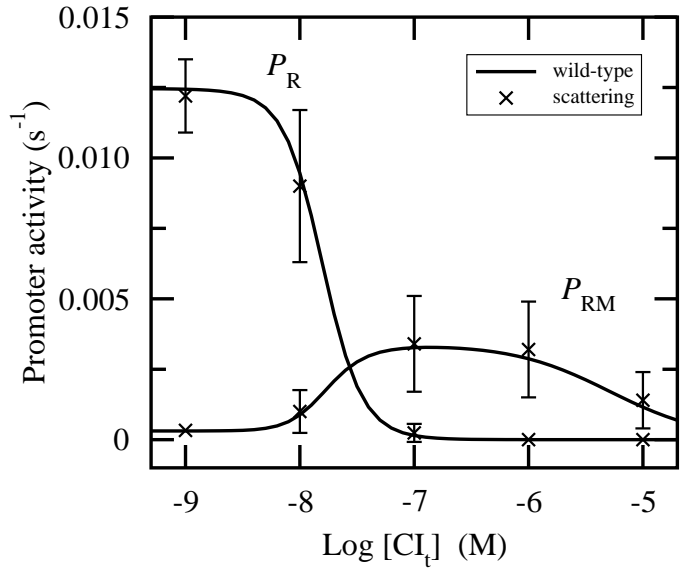,width=10cm}}
   \caption{\label{fig:3} Promoter activity versus total CI concentration for [Cro$_{\rm t}]$ $\approx 50$ nM. See also figure caption of Figure \ref{fig:2}.}
\end{figure}

In order to study the sensitivity of the activity around induction, i.e., at concentrations where CI production is replaced by Cro production, we perform an analogous scattered computation as in Figure \ref{fig:2}, but this time the total Cro concentration ([Cro$_{\rm t}]$) is 50 nM. The latter value may represent a typical Cro concentration around induction\cite{Shea:85,Bakk:03}. Compared with [Cro$_{\rm t}]\approx 0$ the sensitivity of the activity is higher in this concentration regime (see Figure \ref{fig:3}). 
Accordingly, the activities of both $P_{\rm RM}$ and $P_{\rm R}$ are also reduced, which is reasonable because an increased Cro concentration implies increased Cro occupancy at both promoters and transcription occurs less frequently. We also test the case [Cro$_{\rm t}]\approx 200$ nM (typical lytic concentration), that leads to smaller activity than the two previous cases. Due to the small activities, the sensitivity is high in this case.

\begin{figure}
   \centerline{\epsfig{figure=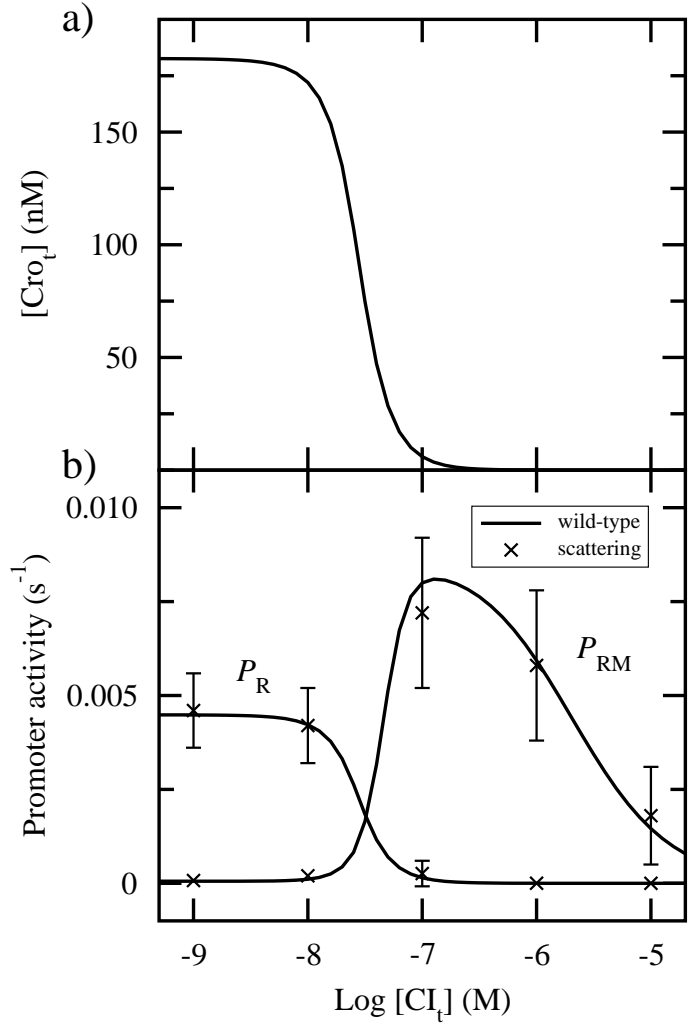,width=10cm}}
   \caption{\label{fig:4}{\bf a)} Total Cro concentration vs.\ total repressor concentration (logarithmic scale) where Cro concentration is determined self-consistently via the equilibrium ansatz $d\,[\text{Cro}_{\rm t}]/dt=0$ in Eq.\ (\ref{eq:Cro_prod}). {\bf b)} Promoter activity versus total CI concentration where Cro concentration is determined self-consistently. Note that the rise of the $P_{\rm RM}$ curve (around 50 nM) is much sharper compared to the situation in Figures \ref{fig:2} and \ref{fig:3}, indicating a larger cooperativity when the Cro concentration is determined in the self-consistent way (feedback).}
\end{figure}

Figures \ref{fig:2} and \ref{fig:3} present the sensitivity of the activity, for a given Cro concentration, versus CI concentration. However, this might be done in a more compact way as  shown in the following. The rate of Cro production may be written as (used to produce Figure \ref{fig:4})\cite{Shea:85,Aurell:02} 
\begin{equation}
  \label{eq:Cro_prod}
\frac{d\,[\text{Cro}_{\rm t}]}{dt}=10^{-9}\,S\,R\,p_{\rm R} -\frac{[\text{Cro}_{\rm t}]}{\tau_{\rm dil}}-\frac{[\text{Cro}_{\rm t}]}{\tau_{\rm deg}},
\end{equation}
 where [Cro$_{\rm t}$] is the total Cro concentration in nM. $S\approx 20$ is the average number of Cro made from each transcript and $R\approx 2.5\times10^{-2}$ s$^{-1}$ is the rate of transcript initiation, both estimated by Aurell et al.\cite{Aurell:02}. $p_{\rm R}$ is the probability of RNAP occupancy of promoter $P_{\rm R}$ calculated from Eq.\ (\ref{eq:f_s}), $\tau_{\rm dil}\approx 34$ min is the life time of a cell generation\cite{Little:99}, and $\tau_{\rm deg}\approx 2600$ s is the {\it in vivo} half-life time of Cro due to degradation\cite{Pakula:86}. The prefactor $10^{-9}$ is simply a conversion factor when going from numbers (of proteins) to concentrations, assuming an average cellular volume of $2\times 10^{-15}$ liters.

We now assume Cro production to be in equilibrium, i.e., $d\,[\text{Cro}_{\rm t}]/dt=0$ in Eq.\ (\ref{eq:Cro_prod}), which is a reasonable assumption because the Cro production occurs on time scale of seconds, while, for instance, a cell generation is of the order of half an hour\cite{Alberts:94}. Thus, for a given repressor concentration we are now able to estimate the Cro concentration (see Figure \ref{fig:4}a). 
One should note that the parameters in Eq.\ (\ref{eq:Cro_prod}) are associated with large uncertainty ($\sim 20\%$), however, this method is a valuable supplement to the presentation in Figures \ref{fig:2} and \ref{fig:3}.

Above we investigated the sensitivity of the activity of the promoters by assuming a fixed Cro concentration (Figures \ref{fig:2} and \ref{fig:3}). In Figure \ref{fig:4}b we show the sensitivity of the activity of the promoters by applying the self-consistent method that corresponds to Figure \ref{fig:4}a. The activity at $P_{\rm R}$ is reduced for [CI$_{\rm t}]<10$ nM, compared with the situation in Figures \ref{fig:2} and \ref{fig:3}. This makes sense, because due to Figure \ref{fig:4}a [Cro$_{\rm t}]\approx 150$ nM for [CI$_{\rm t}]<10$ nM resulting in a self-repression of Cro. $P_{\rm RM}$ is also repressed by Cro in this concentration regime, leading to a zero activity. We find that the sensitivity of the activity is at the same level as in the previous analysis, with a standard deviation of the activity relative to the wild-type activity $>20\%$. 

Finally, we implement the computational scattering method with a flat distribution in an interval $\pm 1.5\times$SD, where SD is the standard deviation in  Table \ref{tab:1}, which corresponds to 67\% confidence intervals. E.g., $\Delta G_2$ is drawn at random in the interval from $-10.8$ kcal/mol to $-10.2$ kcal/mol. This results in a mean value of the activity similar to the wild-type and a sensitivity of the same order as obtained in the Gaussian scattering presented. Thus, the random scattering method seems to be rather insensitive to the functional form of the random drawing distribution function.

\section*{Summary and conclusion}

The main purpose of this work was to study the sensitivity of the production rates (activities) of the regulatory proteins CI and Cro associated with $O_{\rm R}$ (a genetic switch) in phage $\lambda$. The bindings of these regulatory proteins and RNA polymerase to DNA are assumed to be in equilibrium. Thus, by applying a grand canonical approach (statistical open system as presented in Eq.\ (\ref{eq:f_s}))\cite{Hill:60,Ackers:82} we are able to find the probability of binding to  $O_{\rm R}$, whereupon we calculate the rates of CI and Cro production (activities). We perform the computational scattering during which each of the 13 different protein-DNA binding energies are randomly drawn from a Gaussian distribution with mean equivalent to wild-type GFE and standard deviation corresponding to experimental error. Then, the corresponding activities associated with promoters $P_{\rm RM}$ and $P_{\rm R}$ are calculated. This is performed $10^3$ times, whereupon the mean and standard deviation of the resulting activities are evaluated. 

The mean value emerging from  this computational scattering scheme is in general close to wild-type activity, where the latter is calculated from the experimentally (wild-type) given values. The relative sensitivity of the activity, defined as the ratio between the standard deviation ensuing the ``scattering'' and wild-type (unperturbed) activity, is in most cases $>20$\%. The sensitivity of the $P_{\rm RM}$ activity for a lysogen, where  CI concentration
typically is around 200 nM while Cro concentration is zero, is around 20\%. Thus, according to Bailone et al.\cite{Bailone:79}, perturbations of the activities of the size as performed in this work (0.1-0.5 kcal/mol) are not enough to destabilize a lysogen. The $P_{\rm R}$ activity for a lysogen is highly sensitive, however one should note that wild-type activity of  $P_{\rm R}$ is here negligible. Around induction, where both CI and Cro concentrations are at comparable levels (25-50 nM) the sensitivity of the activity is high ($>50\%$). The latter is also the case in the lytic regime where Cro is dominating. Despite the relatively large error, the activities of the two promoters seem to be separated within the error (see Figures \ref{fig:2}, \ref{fig:3}, and \ref{fig:4}) making the switch feasible.   

We note that the perturbations performed here (and conclusions) may to some extent take into account cell-to-cell variations of the concentrations of the proteins, i.e., noise, which effectively may be viewed as variations in the binding energies. However, in order to study noise systematically one should, in the same manner as we scattered GFEs randomly, choose the protein concentrations at random.\cite{Metzler:01,Aurell:02}   

We also make an equilibrium ansatz for Cro production, by which we are able to calculate, for a given Cro concentration, the corresponding repressor concentration. This method leads to a more ``compact'' presentation of data, because then only the CI concentration is a real variable due to the fact that the Cro concentration is implicitly given, or {\it vice versa}. The sensitivity of the activity of the two promoters, due to the latter method, is of the same size as we previously obtained in this work with fixed Cro concentrations.

\section*{Acknowledgments}
K.S.\ acknowledges discussions with I.\ Dodd. R.M.\ would like to thank Joshua Jortner and the organizing committee, Stephen Berry, Ori Cheshnovsky, Joseph Klafter, Abraham Nitzan, and Stuart Rice for the invitation, and an interesting and illustrious JortnerFest.


\end{document}